\documentclass[aps,floats,epsf,twocolumn]{revtex4}
 \usepackage{graphics}

\begin{document}

\title{Quantum critical scaling in magnetic field near the Dirac point in graphene}
\author{Igor F. Herbut and Bitan Roy}

\affiliation{Department of Physics, Simon Fraser University,
 Burnaby, British Columbia, Canada V5A 1S6}

\begin{abstract} Motivated by the recent measurement of the activation energy at the quantum Hall state at
the filling factor $f=1$ in graphene we discuss the scaling of the interaction-induced gaps in vicinity of the Dirac point with the magnetic field. The gap at $f=1$ is shown to be bounded from above by $E(1)/C$, where $E(n)=v_F \sqrt{2nB}$ is the Landau level energy and $C = 5.985 +O(1/N)$ is a universal number.  The universal scaling functions are computed exactly for a large number of Dirac fermions $N$. We find a sublinear dependence of the gap at the laboratory fields of $10 T<B<50T$ for realistic values of short-range repulsion between electrons, and in quantitative agreement with observation.

\end{abstract}
\maketitle

\vspace{10pt}

\section{Introduction}

When placed in a magnetic field, graphene exhibits a series of incompressible quantum Hall states
at filing factors $f=\pm(4n+2)$, which are a direct consequence of the Dirac nature of its' quasiparticles \cite{novoselov}, \cite{zhang}. Whereas this main sequence of states can be understood
within a picture of essentially non-interacting electrons \cite{ando}, \cite{sharapov}, \cite{guinea} the additional  incompressible states at other even filling factors, $f=0$, and at $f=\pm 1$, which become discernible at higher magnetic fields \cite{zhang1}, call for additional considerations.
The observation of the incompressible state at $f=\pm 1$, in particular, implies a complete removal of the fourfold spin and valley degeneracy near the Dirac point, and has been hypothesized to be due to electron-electron interactions. Several theories of these additional incompressible states in graphene have recently been put forward \cite{gusynin, nomura, goerbig,herbut1,fuchs, yang} .

   The idea of an interaction-induced gap at $f=1$ has very recently been given strong support by the experiment of Jiang et al. \cite{jiang} which showed the activation gap in longitudinal resistivity to be independent of the component of the magnetic field parallel to the graphene's plane. Furthermore, gap's dependence on the orthogonal component of the field appears to be sublinear. Both features stand in stark contrast to the behavior at
$f=4$, for example, where the gap scales linearly, and  with the  total magnetic field, which makes it very likely to be due to the
Zeeman effect \cite{zhang1}, \cite{jiang}. The sublinear field dependence of the gap at $f=1$ in particular conforms to the expectation that the interaction-induced gap should scale with the Coulomb energy scale, $e^2 / \epsilon l$, where $l=1/\sqrt{B}$ is the magnetic length.  This natural interpretation, however, runs into the following difficulty: with the commonly assumed dielectric constant $\epsilon\approx 5$ the measured gap at $f=1$ (of $\sim 100K$) is more than an order of magnitude smaller than the theoretical prediction, which would require it to be of a similar size as the energy of the first Landau level. The observed energy gap at $f=1$ provides therefore an unanticipated  intermediate energy scale, in between the Landau level separation ($\sim 1000K$) and the much lower Zeeman energy ($\sim 10K$). The origin of such an energy scale in graphene in the magnetic field  is presently unknown. The purpose of the present paper is to draw attention to this puzzle and propose a solution.

  \begin{figure}[t]
{\centering\resizebox*{85mm}{!}{\includegraphics{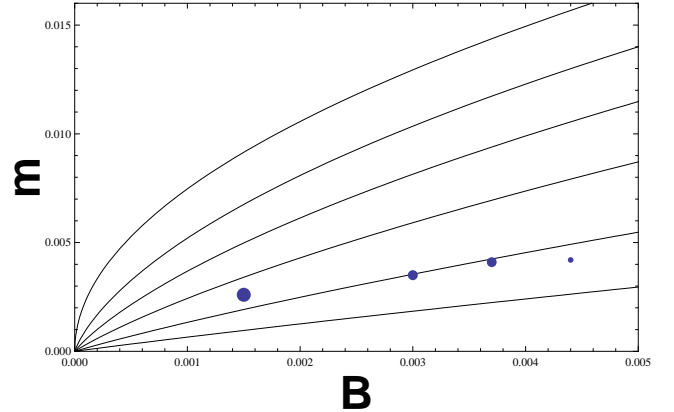}}
\par} \caption[] { The gap $m$ (in units of $v_F \Lambda$) at filling factor $f=1$ as a function of the magnetic field $B$ (in units of $B_0= \Lambda^2 $) below the zero-field critical value of the short-range coupling $g$, for $N=\infty$. The top curve corresponds to the critical point $\delta=0$ (Eq. (1)), and the remaining ones to $\delta=0.03,0.07,0.14, 0.31, 0.7$ (top to bottom), with $\delta=(g_c - g)/g g_c \Lambda $. The best fit to the experimental data \cite{jiang} (dots) assuming $1/\Lambda= 2.5 \AA$ (i. e. $B_0=10^4 T$) is for $\delta=0.31$.}
\end{figure}

  A possible reason for the smallness of the energy gap at $f=1$ may be almost trivial: assuming an order of magnitude larger dielectric constant, which may be due to an accumulated layer of water for example \cite{schedin}, would obviously bring the theory and the observation closer together. Here we wish to put forward an alternative and more general explanation which relies only on the short-range effects of the electron-electron repulsion, and as such completely avoids the ambiguities in the size of graphene's effective dielectric constant. The gist of our theory is the following. A purely short-range repulsion, if above a critical value, would open a gap in graphene even at zero magnetic field \cite{herbut2}. Such a metal-insulator quantum phase transition is believed to be continuous. Right at the metal-insulator quantum critical point then the system becomes scale invariant which, as will be shown, implies that upon the introduction of the magnetic field the gap at $f=1$ behaves as
  \begin{equation}
  m= \frac{v_F \sqrt{2B}}{C}.
  \end{equation}
  $v_F$ is the Fermi velocity at the Dirac point, $\sqrt{2B}$ the energy of the first Landau level in units $\hbar=e/c=1$, and $C$ is a  universal number. We determine the universal number in Eq. (1) to be
  \begin{equation}
  C= 5.985 +O(1/N),
  \end{equation}
  with $N$ as the number of Dirac fermions. Since $1/N$ corrections to simpler universal quantities in the problem, such as the critical exponents at zero magnetic field, are known to be only a few percent even for $N$ as low as two \cite{herbut2}, \cite{wetterich} we expect the above number to be quite indicative of the real value of $C$. At a subcritical value of the short-range interactions, likely to be realized in real graphene, the sublinear dependence $m\sim \sqrt{B}$ crosses over to the linear dependence  $m\sim B$  at lower fields,
   with the values of the gap at all magnetic fields bounded from above by the critical curve in Eq. (1). We compute the universal scaling function for the gap dependence on the magnetic field and the deviation from the criticality in the large-N limit. The family of resulting curves is depicted in Fig. 1. The question posed by the experiment is then whether there exists a line with enough curvature within the range of the laboratory fields which also agrees with the observed magnitude of the gaps. In Fig. 1 we argue that the answer is yes. Our conclusion is strengthened  further by the inclusion of the long-range part of the Coulomb interaction, which provides the leading (logarithmic) corrections to scaling at the large-N critical point. We find its effect to be an additional curvature to the  universal function, and its inclusion thus always improves the comparison with the available experimental data ( see Fig. 3).

      A necessary condition for the applicability of our theory is that the external magnetic field is low compared to
      the characteristic lattice magnetic field scale, or equivalently, that the magnetic length is much longer than the lattice constant. Even at the highest fields of $\sim 45T$ this condition is comfortably satisfied. Provided that the system is not too far in the coupling space from its' quantum critical point,
      in the next section we show that in general under this condition a single relevant coupling constant needs to be included, and a simple scaling law in the magnetic field emerges. In the section III we explicitly compute the scaling function in the large-N limit of the theory, and show how in spite of the complexity of the self-consistent equation that determines the gap, this is possible to do essentially analytically.  In this section we also show how the ultraviolet (UV) cutoff drops out of the universal scaling function for the long magnetic length. In sec. IV leading corrections to scaling due to the long-range tail of the Coulomb interaction are included. Finally, we discuss various aspects of our results in the  concluding sections.

  \section{Scaling}

  Let us begin providing the details behind the above results. To be specific, we assume the spin degeneracy to be removed by the Zeeman splitting, and that the chemical potential is close enough to the Zeeman-shifted Dirac point so that $f=1$. Assuming further a simple quantum critical point at $B=0$ we retain a single relevant short-range coupling constant $g$. We will also include the coupling representing the $\sim 1/r$ tail of the Coulomb interaction, which remains unscreened in graphene: $\lambda= 2 \pi e^2 / \epsilon v_F $ \cite{remark}. At "weak" magnetic fields, at which $l/a \gg 1$, one is at liberty to use the continuum field-theoretic description, with the underlying lattice entering only at energies above the ultraviolet cutoff $\Lambda= 1/a$. The internal consistency of such a description requires that if another value of the cutoff, say $\Lambda/b$, is chosen, the gap in the spectrum $m$ satisfies
    \begin{equation}
    m/\Lambda = b^{-1} v_F (b) F_\pm ( |\delta(b)|, \lambda(b), b^2 B),
    \end{equation}
    where the functions $v_F (b)$, $\delta(b)$ and $\lambda(b)$ are such that the measurable value of $m$ is independent of the arbitrary factor $b$. Here we defined a dimensionless parameter $\delta= (g\Lambda)^{-1} - (g_c \Lambda)^{-1}$ for later convenience, with $g_c$ as the critical value of the short-range coupling. The (engineering) scaling of the magnetic field in the last equation follows from  gauge-invariance. The two functions $F_+$ and $F_-$ refer to $\delta > 0$ ($g<g_c$) and $\delta < 0$ ($g>g_c$), respectively.

    We may then choose $b=(B_0/B)^{1/2}= l/a$, with $B_0= 1/a^2 $, and write
    \begin{equation}
    m= l^{-1} v_F (l/a)  F_\pm [\delta(l/a), \lambda(l/a), B_0].
    \end{equation}
    If $l/a \gg 1$ the flow of the couplings is essentially determined as at zero magnetic field. Omitting all but the single relevant parameter $\delta$ from the outset becomes justified, as all the other couplings are much smaller at a large (magnetic) length scale, provided that in the coupling space the system was not too far from the critical point. Assuming a simple critical point at $B=0$ near which $|\delta(b)| = |\delta| b^{1/\nu}$, $\lambda(b)=\lambda^*$, and $v_F (b)=v_F b^{1-z}$, where $\nu$ and $z$ are the usual correlation length and dynamical critical exponents, we may finally write
    \begin{equation}
    m/(v_F \Lambda) = (a/l)^z  G_\pm [l/\xi, \lambda^*],
    \end{equation}
    with $\xi = a |\delta|^{-\nu}$, as the correlation length. This is the universal scaling form of the gap at low magnetic fields, dictated by the critical point at $\delta=0$, $\lambda=\lambda^*$, and $B=0$  \cite{herbutz}. The last expression is analogous to the finite-size scaling, with the magnetic length playing the role of the system's size \cite{hjv}.

    Before proceeding with the computation of the universal scaling function, let us consider its' limits first. At the critical point,
    \begin{equation}
    G_- [0, \lambda^*]=G_+ [0,\lambda^*].
    \end{equation}
    In particular, for $z=1$ \cite{herbut2} this yields the announced Eq. (1). At $x \gg 1$, we must have
     \begin{equation}
    G_+[x,\lambda^*] \sim x^{z-2},
    \end{equation}
    so that for $g<g_c$ one finds $m\sim B$, as appropriate to "magnetic catalysis" \cite{miransky}. For $g>g_c $, on the other hand, at $x \gg 1$,
    \begin{equation}
    G_- [x, \lambda^*] \sim x^z,
    \end{equation}
    which yields  $m\sim |\delta|^{z\nu}$ above the critical coupling and at $B=0$. This is the familiar scaling of the gap near a quantum critical point \cite{book}.

    \section{Large-N calculation} Let us now take $\lambda=0$, and compute the scaling function $G_+ (x,0)$ in the limit of large number of Dirac fermions. The discussion of the effects of long-range interaction will be presented in the next section. The ground state energy for $N$ species of four-component Dirac fermions in magnetic field and  when $N\rightarrow \infty$ was derived previously in \cite{herbut1}:
    \begin{eqnarray}
    \frac{E(m)-E(0)}{N}= \frac{m^2}{4g} +\frac{B}{4\pi^{3/2}} \int_0 ^\infty \frac{ds}{s^{3/2}} (e^{-s m^2}-1)\\ \nonumber
    [p + 2 K (s\Lambda^2) (\coth(s B)-1)].
    \end{eqnarray}
    $m$ is the gap to be determined by the minimization of $E(m)$, and $g$ is the dominant among the quartic couplings that represent the short-range part of Coulomb repulsion \cite{herbut2}. $K(x)$ is the cutoff function introduced to sum over $n\neq 0$ Landau levels, which satisfies $K(x\rightarrow \infty)=1$ and $K(x\rightarrow 0)=0$, but is otherwise arbitrary. The parameter $p=2$ for $f=0$, when the zeroth Landau level of each Dirac fermion contributes to the energy difference $E(m)-E(0)$, and $p=1$ for $f=N/2$, when half of the zeroth Landau levels do not \cite{herbut1}. $E(m)$ may also be understood as the variational ground state energy of electrons with an on-site, or nearest-neighbor, repulsion on a honeycomb lattice and in low magnetic field, in Hartree approximation.

 Hereafter we set $p=1$, which would correspond to the filling factor $f=1$ for the physical case of $N=2$. Minimizing $E(m)$, after some transformations the gap equation can be written in a compact form as
  \begin{equation}
  y = f(y) + 2 \delta (y B_0/B)^{1/2},
  \end{equation}
  where the variable $y=B/m^2$, and $\delta$ is as defined below Eq. (3), with
  \begin{equation}
  \frac{1}{g_c}=  \frac{\Lambda}{\sqrt{\pi}} \int_0 ^\infty dt \frac{K(t)}{t^{3/2}}.
  \end{equation}
  The function $f(y)$ is defined as
  \begin{equation}
  f(y) = \frac{2}{\sqrt{\pi}} \int_0 ^\infty \frac{dt}{t^{3/2}}  K(\frac{ty B_0}{B} )[1- \frac{ 2 y t e^{-t}}{e^{2 y t }-1} ].
  \end{equation}
  The ultraviolet cutoff enters the gap equation by providing the scale for the magnetic field ($B_0$). In the limit of weak magnetic field, $B/B_0 \ll 1$, we may replace the cutoff function $K(x)$ in the last equation with unity. In this (continuum) limit the dimensionless variable $y$ becomes a {\it universal} function of the ratio
   $\delta(B_0/B)^{1/2}$. The only remaining dependence on the cutoff then is in the value of the critical point $g_c$ which is therefore, as usual, non-universal.

   The gap equation may be solved essentially analytically for $\delta >0 $ by noticing first that at $\delta=0$ the (numerical) solution is at $y_0= 17.913$. This immediately yields the result in Eqs. (1) and (2), with $C = \sqrt{ 2 y_0}$. Since at $\delta>0$ the solution will  be at $y>y_0$, one in fact needs the function $f(y)$ only for large arguments, where it can be expanded as
    \begin{equation}
    f(y) = u y^{1/2} + v y^{-1/2} + O( y^{-3/2} ),
    \end{equation}
    with $u= 4.13031$ and $v=1.84723$.  The difference between $f(y)$ and the first two terms in the expansion on the
    right-hand side is less than a percent already for $y>3$, and decreases further with $y$.  Comparing with the general form in Eq. (5) we may then write the Eq. (10) in terms of the universal function $G_+(x,y)$ as
    \begin{equation}
     -1 + (u + 2 x) G_+ (x,0) + v G_+ ^3 (x, 0)=0,
    \end{equation}
    with the higher order terms in $G_+$ entirely negligible. This way we find $G_+(0,0)= 0.236$, and $G_+(x,0) = 1/2x$ for $x\gg 1$, and uniformly decreasing in between. In sum, at $N=\infty$ and $\lambda=0$ the Eq. (5) becomes
    \begin{equation}
    m= \frac{v_F}{l^z} G_+( \frac{l \delta^\nu}{a},0)
    \end{equation}
    with $z=1$ and $\nu=1+O(1/N)$, and the function $G_+(x,0)$ depicted in Fig. 2.

     \begin{figure}[t]
{\centering\resizebox*{85mm}{!}{\includegraphics{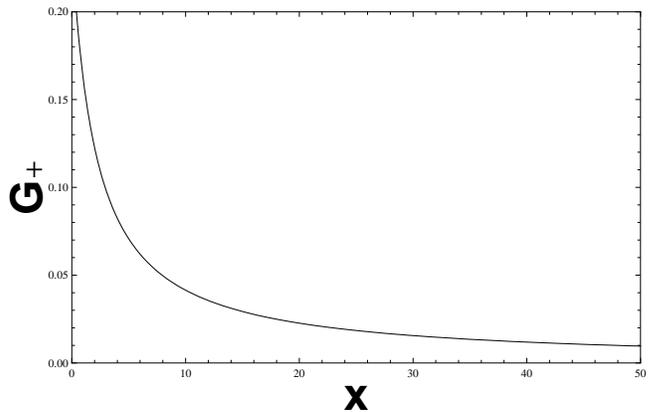}}
\par} \caption[] {The universal scaling function $G_+ (x,0)$ at $N=\infty$.}
\end{figure}

    Although it may not be directly relevant to graphene, for completeness we also briefly describe some of the results when $\delta<0$. Since the gap is finite even at zero field above the critical coupling, the solution of Eq. (10) now lies at $y < y_0$. For weak fields, the relevant regime is $y\ll 1$, where $ f(y)= 4+2y + O(y^2)$. This implies, for example, $G_- (x,0) \rightarrow x/2$ when $x\gg 1$, in agreement with the Eq. (8).  We thus find $m= |\delta|/2 +O(\delta^2)$ in the zero-field limit, and at $N=\infty$.

\section{Leading corrections to scaling}

Let us now turn on a weak long-range interaction $\lambda\ll 1$. It represents a marginally irrelevant perturbation at the large-N critical point $\delta=0$, $\lambda^*=0$ \cite{herbut2}, and thus provides logarithmic corrections to the scaling law in Eq. (5). At $b\gg 1$, the Coulomb interaction scales as
\begin{equation}
\lambda(b)=\frac{8\pi} {\ln b} +O((\frac{1}{\ln b})^2).
\end{equation}
The effect of the long-range Coulomb interaction is to break the pseudo-relativistic invariance of the problem and renormalize the velocity as \cite{gonzales}, \cite{vafekcase}
\begin{equation}
v_F (b) = v_F (1+( \frac{\lambda}{8\pi} +  O(\lambda^2))\ln b).
\end{equation}
The Eq. (4) then yields, for  $l/a \gg 1$
\begin{equation}
m= (\frac{\lambda}{8\pi}+O(\lambda^2)) v_F \frac{\ln(l/a)}{l} G_+( \frac{l\delta^\nu }{a},0).
\end{equation}
Interestingly, the inclusion of such corrections improves the agreement with experiment. As an illustration, in Fig. 3 we display the fit to
\begin{equation}
m=  v_F \sqrt{B} ( 1+ \frac{\lambda}{16 \pi } \ln \frac{B^*}{B} )   G_+( \frac{l\delta^\nu }{a},0).
\end{equation}
$1/\sqrt{B^*}$ defines the length scale appropriate to the measured value of  $v_F=10^6 m/s$, which we will treat as a fitting parameter. The best fit for $\epsilon=1$, for example, is obtained for $B^* = 29 T$. More data seems needed, however, to distinguish between this and the alternative forms such as  a simple $m\sim \sqrt{B}$ considered in ref. 13. We hope our theory will stimulate more experimental work in this direction.

  \begin{figure}[t]
{\centering\resizebox*{85mm}{!}{\includegraphics{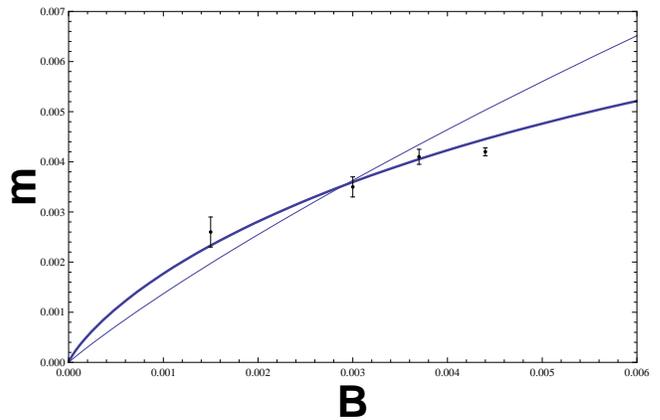}}
\par} \caption[] { The same as in Fig. 1 at $\delta=0.3$, without (dashed line) and with ($\lambda/16\pi= 0.27$, ($\epsilon=1$), thick line) the logarithmic corrections to scaling in Eq. (19).}
\end{figure}

\section{Discussion}

We have argued that taking into account only the short-range parts of the repulsive electron-electron interaction in graphene  suffices to provide a qualitative, and even semi-quantitative understanding of the observed magnetic field dependence of the gap at $f=1$.  Our main assumption is that the value of the dominant short-range coupling is below but not too far from its critical value, so that the aforementioned magnetic field dependence resembles itself as right at the criticality at the laboratory fields.

Note that the precise nature of the short-range coupling, i.e. whether it is on-site or nearest-neighbor interaction, for example, does not matter for the form of the scaling function in the large-N limit we considered. It is only the resulting order parameter \cite{herbut1} that will depend on this.  Any order parameter which breaks the chiral symmetry of the Dirac Hamiltonian and lifts the valley degeneracy in the magnetic field will lead to the same large-N scaling function as obtained here. $1/N$-corrections may depend on the type of the order parameter, however.

  The best fit to the experimental data is provided by $\delta = 0.31$ (Fig. 1), which places the system relatively far from the critical point. On the other hand, assuming the simplest Hubbard model for short-range interactions, for example, would place the critical point at $U_c/t= 4-5$ \cite{herbut2}. With $t=2.5 eV$, the usual estimate \cite{wagner} $U\approx 10 eV$ is then in a reasonable agreement with the value of $\delta$ obtained from the fit. It is in fact not uncommon that a simple single-parameter scaling works well even reasonably far from the critical point. A celebrated example is provided by the scaling plot for classical liquids \cite{book}, where very good scaling of the data is found even at $T/T_c \approx 0.5$.

For weak Zeeman splitting the gap at $f=0$ will also obey the critical scaling in magnetic field, and the critical function can be found in the large-N limit by choosing $p=2$ in the Eq. (9). At $f=0$, however, the Zeeman gap is in competition with the "mass"-gap studied here, and there are reasons to believe that the gap at $f=0$ in reality may be a pure (albeit interaction enhanced) Zeeman gap \cite{herbut1}. For this reason we have not discussed here the scaling at $f=0$ in further detail.

   A test of relevancy of the presented theory would be a measurement of the field-dependent gap at the filling factor $f=1$ on a suspended graphene, for example, where the dielectric constant $\epsilon\approx 1$. If the gaps are still of the same size as in ref. 13 this would eliminate the alternative explanation we mentioned in the introduction, that the smallness of the gap is caused by the extra screening by the surrounding medium. Alternatively, one may want to enhance screening by bringing a metallic plate to the graphene layer. If the gap is indeed caused primarily by the short-range electron-electron interaction as discussed here, this should not alter its size much either.

\section{Conclusion}

We have discussed the general magnetic-field scaling of the gap at filling factor $f=1$ for electrons interacting via short-range repulsive interactions in graphene. Scaling functions and the critical exponents are given explicitly in the large-N limit of the theory, and the leading corrections to scaling deriving from the long-range tail of the Coulomb interaction are determined.  Our theory reproduces the recent experimental data for reasonable values of short-range interactions.

\section{ Acknowledgement}

This work was supported by the NSERC of Canada. We are also grateful to V. Juri\v ci\' c for the critical reading of the manuscript.


\begin{thebibliography}{99}
\bibitem{remark} Such a long-range coupling is also irrelevant at the metal-insulator quantum critical point to the leading order in $1/N$, but only marginally so.
\end{thebibliography}
\end{document}